\PassOptionsToPackage{unicode}{hyperref}
\PassOptionsToPackage{hyphens}{url}
\PassOptionsToPackage{dvipsnames,svgnames,x11names}{xcolor}
\documentclass[
]{article}
\usepackage{amsmath,amssymb}
\usepackage{lmodern}
\usepackage{iftex}
\ifPDFTeX
  \usepackage[T1]{fontenc}
  \usepackage[utf8]{inputenc}
  \usepackage{textcomp} 
\else 
  \usepackage{unicode-math}
  \defaultfontfeatures{Scale=MatchLowercase}
  \defaultfontfeatures[\rmfamily]{Ligatures=TeX,Scale=1}
\fi
\IfFileExists{upquote.sty}{\usepackage{upquote}}{}
\IfFileExists{microtype.sty}{
  \usepackage[]{microtype}
  \UseMicrotypeSet[protrusion]{basicmath} 
}{}
\makeatletter
\@ifundefined{KOMAClassName}{
  \IfFileExists{parskip.sty}{%
    \usepackage{parskip}
  }{
    \setlength{\parindent}{0pt}
    \setlength{\parskip}{6pt plus 2pt minus 1pt}}
}{
  \KOMAoptions{parskip=half}}
\makeatother
\usepackage{xcolor}
\providecommand{\tightlist}{%
  \setlength{\itemsep}{0pt}\setlength{\parskip}{0pt}}
\NewDocumentCommand\citeproctext{}{}
\NewDocumentCommand\citeproc{mm}{%
  \begingroup\def\citeproctext{#2}\cite{#1}\endgroup}
\makeatletter
 \let\@cite@ofmt\@firstofone
 \def\@biblabel#1{}
 \def\@cite#1#2{{#1\if@tempswa , #2\fi}}
\makeatother
\newlength{\cslhangindent}
\newlength{\csllabelwidth}
\newenvironment{CSLReferences}[2] 
 {\begin{list}{}{%
  \setlength{\itemindent}{0pt}
  \setlength{\leftmargin}{0pt}
  \setlength{\parsep}{0pt}
  \ifodd #1
   \setlength{\leftmargin}{\cslhangindent}
   \setlength{\itemindent}{-1\cslhangindent}
  \fi
  \setlength{\itemsep}{#2\baselineskip}}}
 {\end{list}}
\usepackage{calc}

\ifLuaTeX
\usepackage[bidi=basic]{babel}
\else
\usepackage[bidi=default]{babel}
\fi
\babelprovide[main,import]{american}

\def\languageshorthands#1{}
\ifLuaTeX
  \usepackage{selnolig}  
\fi
\IfFileExists{bookmark.sty}{\usepackage{bookmark}}{\usepackage{hyperref}}
\IfFileExists{xurl.sty}{\usepackage{xurl}}{} 
\hypersetup{
  pdftitle={p2smi: A Python Toolkit for Peptide FASTA-to-SMILES
Conversion and Molecular Property Analysis},
  pdfauthor={Aaron L. Feller, Claus O. Wilke},
  pdflang={en-US},
  colorlinks=true,
  linkcolor={Maroon},
  filecolor={Maroon},
  citecolor={Blue},
  urlcolor={Blue},
  pdfcreator={LaTeX via pandoc}}

\title{p2smi: A Python Toolkit for Peptide FASTA-to-SMILES Conversion
and Molecular Property Analysis}

\definecolor{c53baa1}{RGB}{83,186,161}
\definecolor{c202826}{RGB}{32,40,38}

\usepackage[affil-it]{authblk}
\usepackage{orcidlink}
\author[1%
  \ensuremath\mathparagraph]{Aaron L. Feller%
    \,\orcidlink{0000-0002-4476-1026}\,%
    }
\author[1,2%
  ]{Claus O. Wilke%
    \,\orcidlink{0000-0002-7470-9261}\,%
    }

\affil[1]{Department of Interdisciplinary Life Sciences, The University
of Texas at Austin, Austin, TX, United States%
  }
\affil[2]{Department of Integrative Biology, The University of Texas at
Austin, Austin, TX, United States%
  }
\affil[$\mathparagraph$]{Corresponding author. Email: aaron.feller@utexas.edu}

\begin{document}
\maketitle

\section{Summary}\label{summary}

Converting peptide sequences into useful representations for downstream
analysis is a common step in computational modeling and cheminformatics.
Furthermore, peptide drugs (e.g.~Semaglutide, Degarelix) often take
advantage of the diverse chemistries found in noncanonical amino acids
(NCAAs), altered stereochemistry, and backbone modifications. Despite
there being several chemoinformatics toolkits, none are tailored to the
task of converting a modified peptide from an amino acid representation
to the chemical string nomenclature Simplified Molecular-Input
Line-Entry System (SMILES), often used in chemical modeling. Here we
present p2smi, a Python toolkit with CLI, designed to facilitate the
conversion of peptide sequences into chemical SMILES strings. By
supporting both cyclic and linear peptides, including those with NCAAs,
p2smi enables researchers to generate accurate SMILES strings for
drug-like peptides, reducing the overhead for computational modeling and
cheminformatics analyses. The toolkit also offers functionalities for
chemical modification, synthesis feasibility evaluation, and calculation
of molecular properties such as hydrophobicity, topological polar
surface area, molecular weight, and adherence to Lipinski's rules for
drug-likeness.

\section{Statement of need}\label{statement-of-need}

Several general bioinformatics toolkits exist for chemical
representation and cheminformatics workflows
(\citeproc{ref-ChemAxon}{\emph{ChemAxon}, 2025};
\citeproc{ref-landrum2013rdkit}{Landrum, 2025};
\citeproc{ref-o2011open}{O'Boyle et al., 2011};
\citeproc{ref-OEChem}{OpenEye, 2025}); however, many face limitations
such as proprietary licensing and lack of specific functionalities for
drug-like peptides. These constraints limit high-throughput application
of sequence generation and conversion, especially for peptides
incorporating noncanonical amino acids (NCAAs), diverse stereochemistry,
and common chemical modifications. The development of p2smi was driven
by the need to generate large-scale datasets of peptide SMILES strings
for pretraining transformer-based models to understand SMILES notation
(\citeproc{ref-feller2025peptide}{Feller \& Wilke, 2025}). Built on the
core concepts from the CycloPs (\citeproc{ref-duffy2011cyclops}{Duffy et
al., 2011}) method for FASTA-to-SMILES conversion, p2smi has evolved
into a stand-alone resource to support peptide-focused machine learning
pipelines and peptide design workflows. We used p2smi to build a dataset
of 10M peptides with NCAAs, backbone modifications, and cyclizations for
pretraining a chemical language model that was used for predicting
peptide diffusion across an artificial cell membrane
(\citeproc{ref-feller2025peptide}{Feller \& Wilke, 2025}). We have made
p2smi available as a pip-installable package, offering both command-line
tools and Python functions for seamless integration into larger
workflows.

\section{Features}\label{features}

By leveraging the database in SwissSidechain
(\citeproc{ref-gfeller2012swisssidechain}{Gfeller et al., 2012}), p2smi
accommodates over 100 unnatural amino acid residues. Our package
supports multiple cyclization chemistries, including disulfide bonds,
head-to-tail, and side-chain cyclizations. Additionally, p2smi offers a
SMILES modification tool, allowing users to apply N-methylation and
PEGylation---modifications often used to influence peptide-drug
stability and bioactivity. An integrated synthetic feasibility check
assists researchers in assessing the practical synthesis of natural
peptides. Furthermore, p2smi computes key molecular properties such as
logP, TPSA, molecular weight, and Lipinski's rule compliance, supporting
early-stage drug-likeness evaluation. Collectively, these features
position p2smi as a useful tool for both computational peptide modeling
and experimental design.

To install p2smi, use the \texttt{pip\ install\ p2smi} command. Once
installed, p2smi offers five primary command-line tools designed to
facilitate various aspects of peptide analysis and modification:

\begin{itemize}
\tightlist
\item
  generate-peptides: This tool enables the generation of random peptide
  sequences based on user-defined constraints and modifications,
  allowing for the creation of diverse peptide libraries for
  computational studies.
\item
  fasta2smi: Converts peptide sequences from FASTA format into SMILES
  notation, facilitating integration with cheminformatics workflows that
  utilize SMILES strings for molecular representation.
\item
  modify-smiles: Applies specific chemical modifications, such as
  N-methylation and PEGylation, to existing SMILES strings, enabling the
  exploration of modified peptides' properties and behaviors.
\item
  smiles-props: Computes molecular properties---including logP,
  topological polar surface area (TPSA), molecular formula, and
  evaluates compliance with Lipinski's rules---from provided SMILES
  strings, assisting in the assessment of peptides' drug-like
  characteristics.
\item
  synthesis-check: Evaluates the synthetic feasibility of peptides based
  on defined synthesis rules, aiding researchers in determining the
  practicality of synthesizing specific peptide sequences.
\end{itemize}

For detailed usage instructions and options for each command, users can
append the --help flag to any command (e.g., generate-peptides --help).
This will provide guidance on the command's functionality and available
parameters.

\section{State of the field}\label{state-of-the-field}

In the realm of peptide informatics, several tools have been recently
developed to facilitate the analysis and representation of peptides,
particularly those incorporating NCAAs and complex modifications.
Notably, pyPept (\citeproc{ref-ochoa2023pypept}{Ochoa et al., 2023}) and
PepFuNN (\citeproc{ref-ochoa2025pepfunn}{Ochoa \& Deibler, 2025}) have
emerged as significant contributions in this area.

pyPept is a Python library that generates 2D and 3D representations of
peptides. It converts sequences from formats like FASTA, HELM, or BILN
into molecular graphs, enabling visualization and physicochemical
property calculations. Notably, pyPept allows customization of monomer
libraries to accommodate a wide range of peptide modifications. It also
offers modules for rapid peptide conformer generation, incorporating
user-defined or predicted secondary structure restraints, which is
valuable for structural analyses.

PepFuNN is an open-source Python package designed to explore the
chemical space of peptide libraries and conduct structure--activity
relationship analyses. It includes modules for calculating
physicochemical properties, assessing similarity using various peptide
representations, clustering peptides based on molecular fingerprints or
descriptors, and designing peptide libraries tailored to specific
requirements. Additionally, PepFuNN provides tools for extracting
matched pairs from experimental data, aiding in the identification of
key mutations for subsequent design iterations.

While both tools offer valuable capabilities, they are not specifically
designed for the direct conversion of peptide sequences into SMILES
strings---a functionality central to the initial use-case for p2smi of
generating a large-scale database. Rather, pyPept and PepFuNN focus on
structural representation, analysis, and structure--activity
relationship studies of peptides, complementing the sequence-to-SMILES
conversion capabilities provided by p2smi.

\section{Code availability}
We have provided p2smi as a pip-installable package, available on PyPI at https://pypi.org/project/p2smi. The source code, including documentation and example notebooks, is openly available on GitHub at https://github.com/aaronfeller/p2smi.

\section{Acknowledgements}\label{acknowledgements}

This work was supported by NIH grant 1R01 AI148419. C.O.W. was also
supported by the Blumberg Centennial Professorship in Molecular
Evolution at The University of Texas at Austin.

\section*{References}\label{references}
\addcontentsline{toc}{section}{References}

\phantomsection\label{refs}

\end{document}